\documentclass[aps,showpacs,pra,amssymb, amsmath, twocolumn]{revtex4}
\usepackage{graphicx}
\usepackage{amssymb}
\usepackage{amsmath}
\usepackage{bm}
\begin{document}
\title{Single-laser-pulse implementation of arbitrary ZYZ rotations of an atomic qubit}
\author{Han-gyeol Lee, Yunheung Song, and Jaewook Ahn}
\email{jwahn@kaist.ac.kr}
\address{Department of Physics, KAIST, Daejeon 305-701, Korea}

\begin{abstract}
Arbitrary rotation of a qubit can be performed with a three-pulse sequence; for example, ZYZ rotations. However, this requires precise control of the relative phase and timing between the pulses, making it technically challenging in optical implementation in a short time scale. Here we show any ZYZ rotations can be implemented with a single laser-pulse, that is {\it a chirped pulse with a temporal hole}. The hole of this shaped pulse induces a non-adiabatic interaction in the middle of the adiabatic evolution of the chirped pulse, converting the central part of an otherwise simple Z-rotation to a Y rotation, constructing ZYZ rotations. The result of our experiment performed with shaped femtosecond laser pulses and cold rubidium atoms shows strong agreement with the theory.
\end{abstract}

\maketitle

\section{Introduction}
Qubit is the information stored in the quantum state of a two-level system, routinely used as the smallest unit of information processed in the quantum circuit model of quantum computation~\cite{NielsenChuang}. In order to construct a universal computational gate set, single-qubit rotations, about at least two distinct rotational axes are required as well as a two-qubit gate, e.g., the CNOT gate. Single-qubit rotation gates, such as Hadamard and Pauli X, Y, and Z gates have been implemented on numerous physical systems, including photons~\cite{photons}, ions~\cite{ions}, atoms~\cite{atoms}, molecules~\cite{molecules}, quantum dots~\cite{aatoms}, and superconducting qubits~\cite{scqubits}.

Many single-qubit rotations in a sequence can also be performed with {\it a single arbitrary rotation gate}, which simplifies otherwise complex physical implementation of many distinct rotations in a unified fashion. An arbitrary rotation (of rotation angle $\phi$ and rotational axis $\hat{n}$) can be constructed with a minimum of three rotations that correspond to the set of Euler angle rotations: for example, the three rotations in the best-known ZYZ-decomposition are given by 
\begin{equation}
\mathcal{R}_{\hat{n}}(\phi) = \mathcal{R}_{\hat{z}}(\Phi_2)\mathcal{R}_{\hat{y}}(\Theta)\mathcal{R}_{\hat{z}}(\Phi_1),
\end{equation}
where $\mathcal{R}$ represents a rotational transformation, and $\hat{n}$ and $\phi$ are respectively given as a function of three rotation angles $\Phi_1$, $\Phi_2$, and $\Theta$~\cite{arbitrary}. In an optical implementation of two-level system dynamics, Z-rotations use either a time-evolution or a far-detuned excitation~\cite{LimSR2014,Weiss2016}, and X or Y-rotations a resonant area-pulse interaction, both of which and their combinations require a precise control of the relative phase and timing among the constituent pulsed interactions. 

In this paper, we show that an arbitrary rotation can be, alternatively, performed with a single laser-pulse, when the pulse is programmed to be {\it a chirped pulse with a temporal hole}. As to be discussed in the rest of the paper, a single laser pulse with the given pulse shape can implement  ZYZ-decomposed rotations all at once, where the temporal hole in the middle of a chirped pulse induces a strong non-adiabatic evolution, which is a Y-rotation, amid an otherwise monotonic adiabatic evolution, a Z-rotation, due to the chirped pulse. The predicted behavior of the ZYZ-decomposition is to be experimentally verified with cold atomic qubits and as-programmed femtosecond laser pulses.

\section{Theoretical analysis}

\begin{figure*}
\centerline{\includegraphics[width= 0.9 \textwidth]{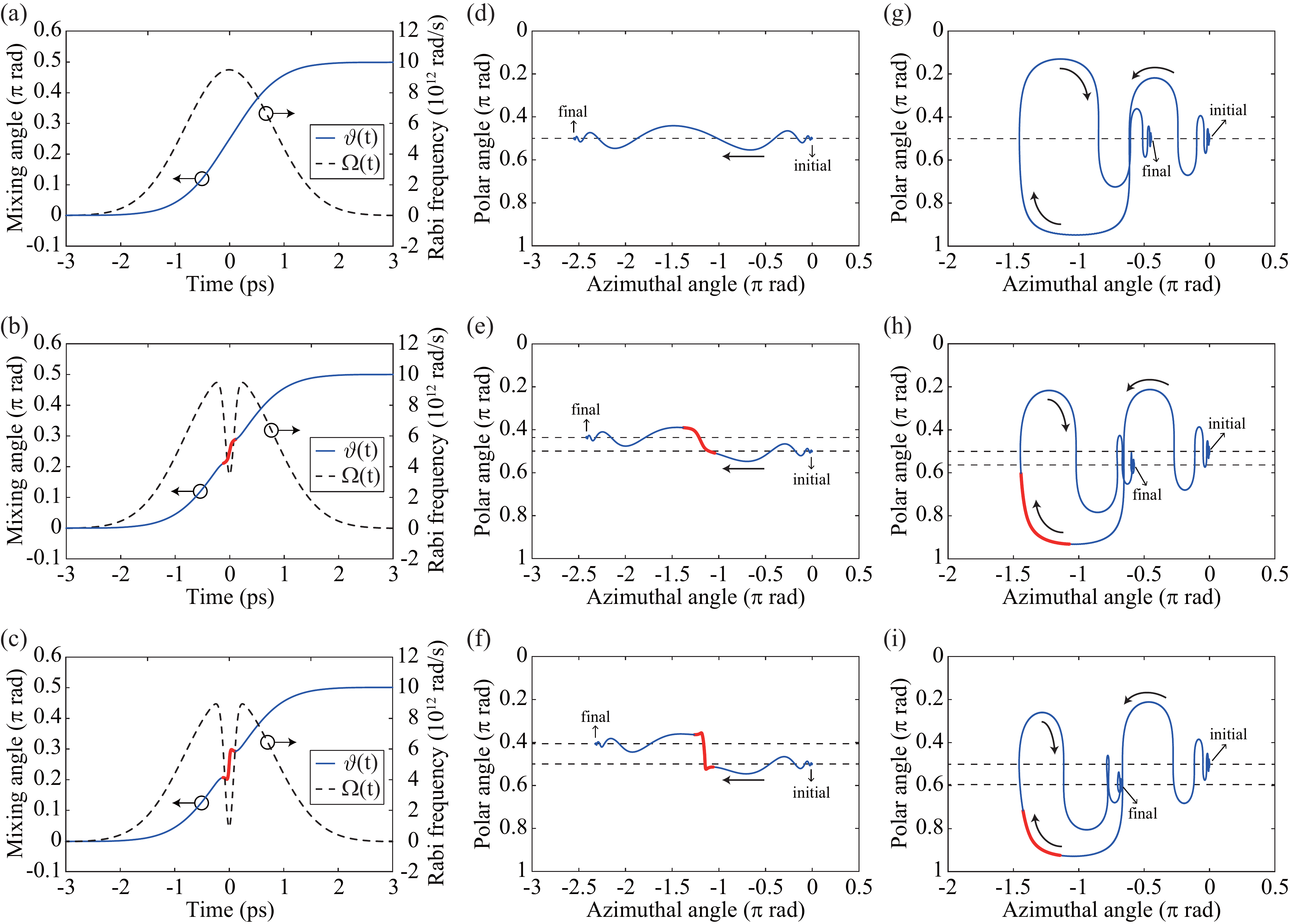}}
    \caption{(Color online) (a-c) Time dependence of Rabi frequency $\Omega$ and mixing angle $\vartheta$ plotted for (a) a chirped pulse, (b) a chirped pulse with a temporal hole of width $\tau_h = 0.1\tau$ and depth $k=0.65$, and (c) of $\tau_h = 0.1\tau$ and $k=9$. (d-f) Bloch vector evolution of the adiabatic states in the ``detuning'' interaction picture, corresponding to (a), (b), and (c), where the $x$-axis is the azimuthal angle of the Bloch sphere and the $y$-axis is the polar angle. In (d-f), the north pole (when polar angle is 0) is $|0\rangle_{\Delta}$ and the south pole (when polar angle is $\pi$) is $|1\rangle_{\Delta}$. (g-i) Bloch vector evolution in the ``atomic'' interaction picture corresponding to (a), (b), and (c).  In (g-i), the north pole (when polar angle is 0) is $|g\rangle_{\omega_0}$ and the south pole (when polar angle is $\pi$) is $|e\rangle_{\omega_0}$. The thick red lines in (b), (c), (e), (f), (h) and (i) indicate the $-\tau_h<t<\tau_h$ region (see text). The horizontal dashed lines in (d-i) indicate the polar angles of the initial and final states, which show that the amounts of change in the polar angle are the same between the ``detuning'' and ``atomic'' interaction pictures.} 
\label{Fig1}
\end{figure*}
 
We consider the dynamics of a two-level atom, driven by a chirped laser pulse with a temporal hole. The electric field of the pulse, where both the main pulse and the hole are assumed to be of Gaussian pulse shape, is given by 
\begin{eqnarray}
E(t) = A_{0}(e^{-{t^2}/{\tau^2}}-ke^{-{t^2}/{\tau_h^2}})\cos(\omega_{0}t+\alpha t^2+\varphi),
\label{efieldhole}
\end{eqnarray}
where $A_0$ is the amplitude, $\tau$ and $\tau_h$ are respectively the widths of the main pulse and the hole, $k$ ($0\leq k\leq 1$) is the depth of the hole, $\alpha$ is the linear chirp parameter, and $\varphi$ is the carrier phase (see Appendix A). The contribution of the carrier phase is a simple Z-rotation, {\it i.e.} $\mathcal{R}_{\hat{z}}(\varphi)$, so we will first consider the $\varphi=0$ case. When the base vectors are defined by $|g\rangle$ and $|e\rangle$ (of respective energies $-\hbar\omega_0 /2$ and $\hbar \omega_0 /2$), the Hamiltonian in the adiabatic basis~\cite{ShoreBook, AllenBook} (see Appendix B), after the rotating wave approximation, is given by
\begin{equation}
\label{adiaH1}
H_A 
= \frac{\hbar}{2} \left[
 \begin{array}{cc}
\lambda_- & -2i\dot{\vartheta}  \\
2i\dot{\vartheta}  & \lambda_+
\end{array} 
\right],
\end{equation}
where $\lambda_\pm = \pm \sqrt{\Omega^2+\Delta^2}$ are the eigenvalues, for  the Rabi frequency $\Omega$ and  the instantaneous detuning $\Delta=-2\alpha t$, and $\vartheta$ is the adiabatic mixing angle defined by $2\vartheta = \tan^{-1} {\Omega}/{\Delta}$ for $0 \le\vartheta \le {\pi}/{2}$. However, with Eq.~\eqref{adiaH1}, the phase of the state diverges at $t\to\pm\infty$, so we use an additional transformation $\mathcal{T}_{\Delta}=\exp\Big({i} \int_{0}^{t} T_\Delta dt'/{\hbar} \Big)$ with $T_\Delta = \frac{\hbar}{2}\left[ \begin{array}{cc}
-|\Delta| & 0  \\
0 & |\Delta|
\end{array} \right]$ to remove this rapidly oscillating phase. The resulting Hamiltonian that represents the dynamics of the adiabatic state in the ``detuning'' interaction picture is given by
\begin{equation}
\label{adiaH2}
H_{\Delta}  = \frac{\hbar}{2} \left[
 \begin{array}{cc}
|\Delta|-\sqrt{\Delta^2+\Omega^2}&-2i\dot{\vartheta}e^{-i|\Delta|/2}  \\
2i\dot{\vartheta}e^{i|\Delta|/2}  & \sqrt{\Delta^2+\Omega^2}-|\Delta|
\end{array} 
\right],
\end{equation}
and corresponding base vectors are $|0(t)\rangle_\Delta$ and $|1(t)\rangle_\Delta$.

Figure~\ref{Fig1} shows the behavior of the mixing angle $\vartheta$, compared with the Rabi frequency $\Omega$ for various hole depth $k$ (first column), and the corresponding Bloch vector evolution in the ``detuning'' interaction picture (second column) and in the ``atomic'' interaction picture (third column).  The pulseand the transformed base vectors are $|g\rangle_{\omega_0} = \mathcal{T}_{\omega_0}|g\rangle$ and $|e\rangle_{\omega_0} = \mathcal{T}_{\omega_0}|e\rangle$. Then, using an arbitrary state 
$\psi(t)$, the relation between the interaction picture of the atomic basis (labeled with $\omega_0$) and the interaction picture of the adiabatic basis (labeled with $\Delta$) is given by
\begin{eqnarray}
|\psi(t)\rangle_{\Delta} = \mathcal{T}_{\Delta}(t)R(\vartheta(t))\mathcal{T}_{\omega_L}(t)\mathcal{T}_{\omega_0}^{\dagger}(t)|\psi(t)\rangle_{\omega_0},
\label{ap_staterel1}
\end{eqnarray} without a hole in Fig.~\ref{Fig1}(a) shows slow change in $\vartheta$ and relatively large $\Omega$, suggesting that the adiabatic condition, $2\dot{\vartheta}\ll |\lambda_+-\lambda_-|$, is satisfied in all time. So, a pulse without a hole induces an adiabatic evolution, {i.e.}, a Z-rotation in the adiabatic basis, as depicted in Fig.~\ref{Fig1}(d).

On the other hand, the pulses with a hole in Figs.~\ref{Fig1}(b) and \ref{Fig1}(c) exhibit abrupt change in $\vartheta$ near $t=0$.
Therefore, the overall dynamics can be decomposed to sub-dynamics in three different time zones: $t<-\tau_h$, $-\tau_h<t<\tau_h$, and $t>\tau_h$, as shown in Figs.~\ref{Fig1}(e) and \ref{Fig1}(f). In the central time zone ($-\tau_h<t<\tau_h$), the hole makes $\Omega$ small and rapid change in $\vartheta$ occurs. Since the Hamiltonian is dominated by the non-adiabatic coupling (the off-diagonal components), it is approximately given by
\begin{equation}
\label{adiaY}
H_{\Delta}(t\approx 0)  
\approx \frac{\hbar}{2} \left[
 \begin{array}{cc}
0 & -2i\dot{\vartheta}  \\
2i\dot{\vartheta}  & 0
\end{array} 
\right],
\end{equation}
which corresponds to the Y-rotation with rotation angle 
\begin{eqnarray}
\label{yrotangle}
\Theta &\approx& \int_{-\tau_h}^{\tau_h} 2\dot{\vartheta} dt 
= 2\left[\vartheta(\tau_h)-\vartheta(-\tau_h)\right].
\end{eqnarray}
In both side regions  ($t<-\tau_h$ and $t>\tau_h$), Z-rotations occur due to the adiabatic evolution of the chirped pulse. The rotation angles are respectively given by
\begin{eqnarray}
\label{zrotangle}
\Phi_1  &\approx &\int_{-\infty}^{-\tau_h} \left[ |\Delta(t)| - \sqrt{\Delta^2(t)+\Omega^2(t)}\right] dt  \\
\Phi_2  &\approx &\int_{\tau_h}^{\infty} \left[ |\Delta(t)| - \sqrt{\Delta^2(t)+\Omega^2(t)}\right] dt,
\end{eqnarray}
and, as a result, the total time-evolution, including the Z-rotation due to the carrier phase $\mathcal{R}_{\hat{z}}(\varphi)$, is given  by  
\begin{eqnarray}
\lefteqn{\mathcal{R}_{\hat{z}}(\Phi_2)\mathcal{R}_{\hat{y}}(\Theta)\mathcal{R}_{\hat{z}}(\Phi_1+\varphi)}    
\nonumber \\
&=&
\left[ \begin{array}{cc}
e^{-i (\Phi_1+\Phi_2+ \varphi)/2} \cos \frac{\Theta}{2} & 
-e^{i\varphi/2} \sin \frac{\Theta}{2}  \\
e^{-i\varphi/2} \sin \frac{\Theta}{2}   & 
e^{i(\varphi+\Phi_1+\Phi_2)/2} \cos \frac{\Theta}{2}
\end{array} \right],
\label{eq.rrr}
\end{eqnarray}
which corresponds to an arbitrary ZYZ rotation with three parameters $\Phi_1+\varphi$, $\Phi_2$, and $\Theta$ that can be made fully independent. 

Although the ZYZ rotation in the Eq.~\eqref{eq.rrr} is derived for the adiabatic states in the ``detuning'' interaction picture, $|\psi(t)\rangle_\Delta=\mathcal{T}_\Delta |\psi(t)\rangle_A$, the result is also valid for the corresponding original atomic states in the ``atomic'' interaction picture, $|\psi(t)\rangle_{\omega_0}= \mathcal{T}_{\omega_0} |\psi(t)\rangle$ (see Appendix B for the definition), because of the simple relation between these two states at $t=\pm\infty$. The relation between these two states are given by
\begin{eqnarray}
|\psi(t)\rangle_{\Delta} = \mathcal{T}_{\Delta}(t)R(\vartheta(t))\mathcal{T}_{\omega_L}(t)\mathcal{T}_{\omega_0}^{\dagger}(t)|\psi(t)\rangle_{\omega_0},
\label{ap_staterel1}
\end{eqnarray}
where the $\mathcal{T}_{\omega_L}$ and $R(\vartheta)$ are the transformation to the ``field'' interaction picture and the adiabatic transform matrix  (see Appendix B for details). At extreme times, $t=\pm \infty$, the overall transformation becomes simple, given by
\begin{equation}
\mathcal{T}_{\Delta}(\pm \infty)R(\vartheta(\pm \infty))\mathcal{T}_{\omega_L}(\pm \infty)\mathcal{T}_{\omega_0}^{\dagger}(\pm \infty)=R(\vartheta(\pm \infty)),
\label{ap_staterel1}
\end{equation}
with $R(\vartheta(-\infty)) = \left(\begin{array}{cc} 1 & 0  \\ 0 & 1 \end{array}\right) $ and $R(\vartheta(\infty)) = \left( \begin{array}{cc} 0 & -1  \\ 1 & 0 \end{array}\right) $.
The base vectors in these two representations are identical ($|0\rangle_\Delta = |g\rangle_{\omega_0}$, $|1\rangle_\Delta = |e\rangle_{\omega_0}$) at $t=-\infty$ and switched ($|0\rangle_\Delta = -|e\rangle_{\omega_0}$, $|1\rangle_\Delta = |g\rangle_{\omega_0}$) at $t=\infty$. Therefore, the time evolution in Eq.~\eqref{eq.rrr}, the ZYZ rotations, defined in the $\{|0\rangle_\Delta, |1\rangle_\Delta\}$ basis (the ``detuning'' interaction picture) can be also written as  
\begin{equation}
R^{\dagger}(\vartheta(\infty))\mathcal{R}_{\hat{z}}(\Phi_2)\mathcal{R}_{\hat{y}}(\Theta)\mathcal{R}_{\hat{z}}(\Phi_1+\varphi)
\label{tevolrel}
\end{equation}
in the $\{|g\rangle_{\omega_0}, |e\rangle_{\omega_0}\}$ basis (the ``atomic'' interaction picture).

The third column in Fig.~\ref{Fig1} shows the corresponding time-evolution in the ``atomic'' interaction picture. The net changes of the state vector between the initial and final states are the same as those in the second column (the ``detuning'' interaction picture). Otherwise complicated time-evolutions of the state vector, e.g., in the ``atomic'' interaction basis, can be easily decomposed to the ZYZ rotations in our ``detuning'' interaction picture.

\begin{figure}
\centerline{\includegraphics[width=0.4\textwidth]{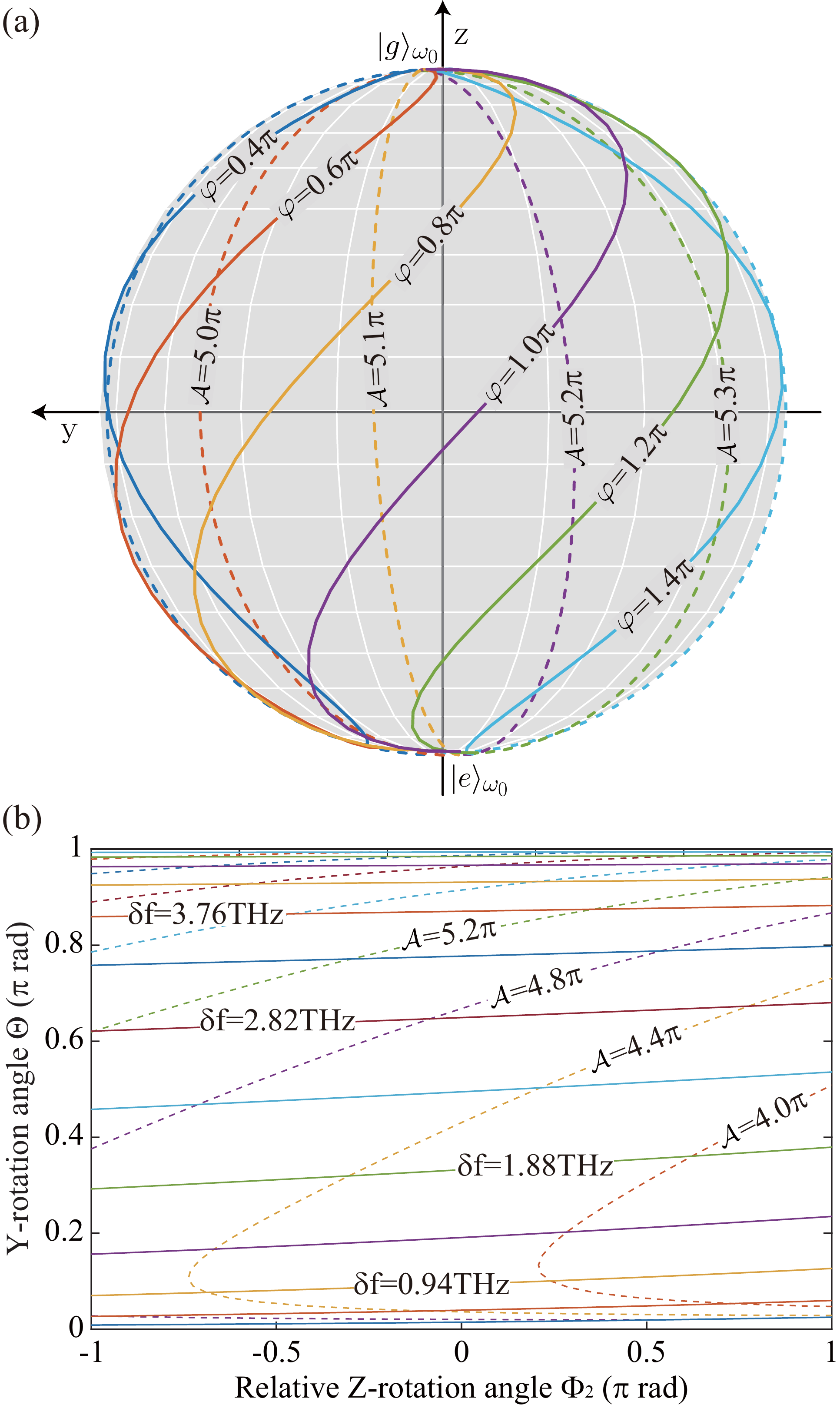}}
    \caption{(Color online) (a) Final states on the Bloch sphere in the ``atomic'' interaction picture, spanned by the resonant chirped pulse with a temporal hole ($\Phi_1=\Phi_2$ case), for an initial state $(|g\rangle_{\omega_0}+|e\rangle_{\omega_0})/\sqrt{2}$.  The carrier phase $\varphi$ and the equivalent TL pulse area $\mathcal{A}$, defined in Eq.~\eqref{ap_pulsearea}, were varied, while $\tau=5.9$~ps, $\alpha=1.25$~rad/ps$^2$, $\tau_h=0.5\tau$, and $k=0.95$ were kept constant.  (b) Detuning $\delta f = 2\alpha\delta t / 2 \pi $, that is associated with the time shift $\delta t$ in Eq.~\eqref{detuning}, was used as an additional control parameter to show that (the second) Z- and Y-rotation angles, $\Phi_2$ and $\Theta$, are fully spanned, where $\tau=2.95$~ps, $\alpha=2.5$~rad/ps$^2$, $\tau_h=0.5\tau$ and $k=0.7$.} 
\label{fig2}
\end{figure}

Figure ~\ref{fig2} demonstrates the arbitrary qubit rotations. The numerical calculation in Fig.~\ref{fig2}(a) shows Bloch sphere points accessible by as-shaped pulses controlled with two parameters $\mathcal{A}$ (the pulse area) and $\varphi$ (the carrier phase).  When the pulse envelope is symmetric as in Eq.~\eqref{efieldhole}, $\Phi_1$ equals $\Phi_2$. In this case and also when the qubit starts from the initial state given by
\begin{eqnarray}
|\psi_{\rm init}\rangle 
&=& \frac{1}{
\sqrt{2}}\left(|0(-\infty)\rangle_{\Delta}+|1(-\infty)\rangle_{\Delta}\right) \nonumber \\
&=& \frac{1}{
\sqrt{2}}\left(|g\rangle_{\omega_0}+|e\rangle_{\omega_0}\right),
\label{eq.init}
\end{eqnarray}
any final positions on the Bloch sphere are accessible, as shown in Fig.~\ref{fig2}(a). Even without assuming such an initial state, full arbitrariness can be achieved with an addition degree of freedom in pulse shaping. When detuning $\delta\omega$ is implemented by a time shift, $\delta t = \delta \omega / 2\alpha$, of the main pulse,  the electric field is given by 
\begin{equation}
E(t) = A_{0}[e^{-{(t-\delta t)^2}/{\tau^2}}(1-ke^{-{t^2}/{\tau_h^2}})]\cos(\omega_{0}t+\alpha t^2),
\label{detuning}
\end{equation}
where the hole is fixed at $t=0$. As shown in Fig.~\ref{fig2}(b), then the full range range $2\pi$ for $\Phi_{2}$ and $\pi$ for $\Theta$ are completely spanned, ensuring the given ZYZ rotations to be arbitrary. 

We note that the equivalent transform-limited pulse area, $\mathcal{A}$ in Fig.~\ref{fig2} is defined with the pulse area of a transform-limited (TL) pulse that has the same pulse-energy with the shaped pulse, which is given by
\begin{eqnarray}
\mathcal{A} &=&\frac{\mu}{\hbar}\int_{-\infty}^{\infty} dt E_0  e^{-{t^2}/{\tau_0^2}} \nonumber \\ &=& \frac{2\mu}{\hbar}\sqrt{\tau_0\sqrt{\frac{\pi}{2}}\int_{-\infty}^{\infty}dt  |E_{{\rm shaped}}(t)|^2}
\label{ap_pulsearea},
\end{eqnarray}
where $\tau_0$ is the pulse width of the TL pulse.
With this definition, the pulse energies of the shaped pulse and the TL pulse are equal {\it i.e.},
\begin{equation}
\int_{-\infty}^{\infty} dt|E_0e^{-t^2/\tau_0^2}\cos(\omega_0 t)|^2 =  \int_{-\infty}^{\infty} dt|E_{\rm shaped}(t)|^2.
\end{equation}

\section{Experimental verification}

\begin{figure}
\centerline{\includegraphics[width=0.4\textwidth]{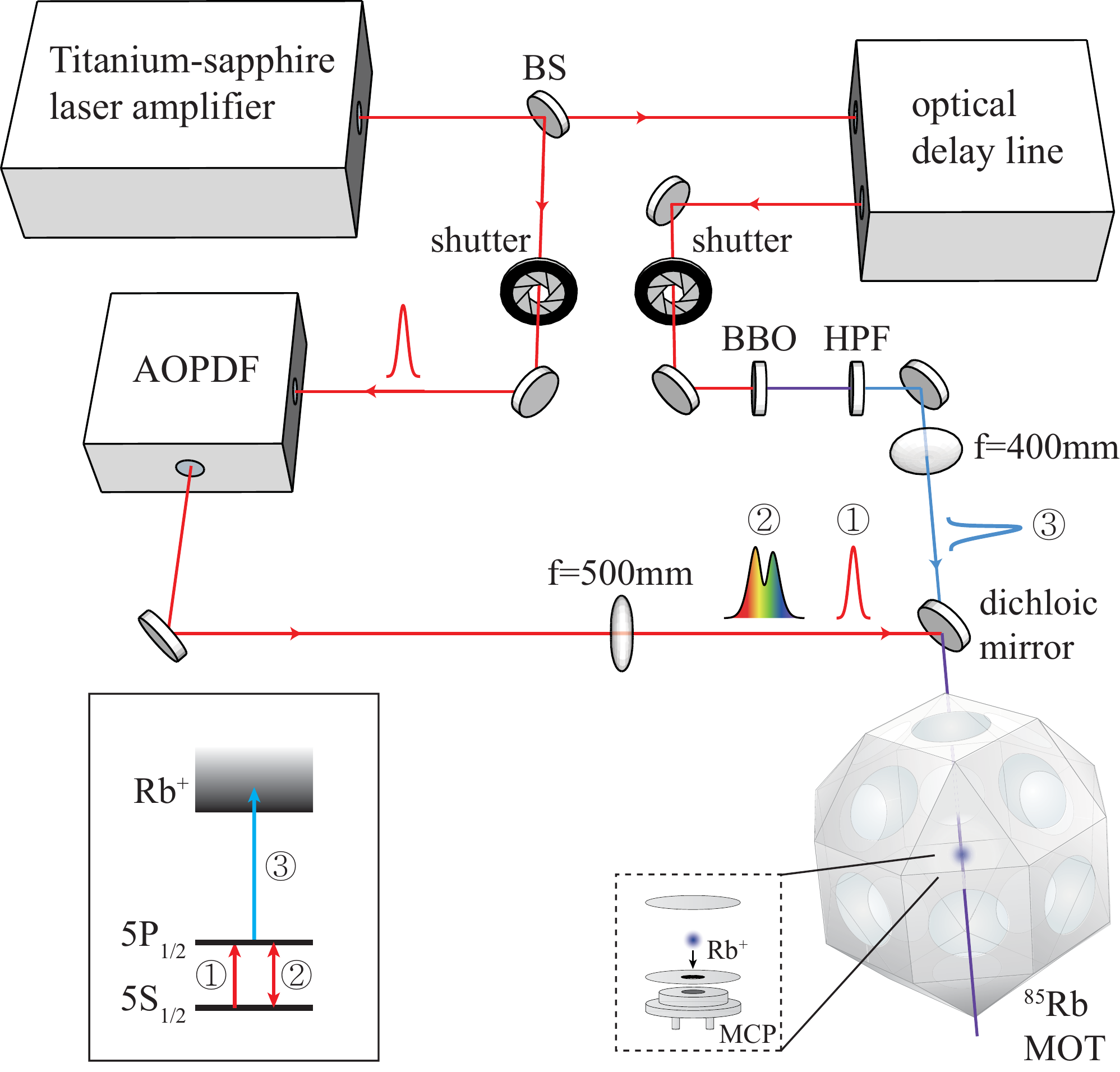}}
    \caption{(Color online) Schematic of the experimental setup: Laser pulses were from the femtosecond laser and programmed to operate arbitrary single qubit rotations of cold atom qubits in a magneto-optical trap. Inset shows the energy level structure of the atomic qubit along with ionization states. The arrows labeled with numbers illustrate the experimental pulse sequence.}
    \label{Fig3}
\end{figure}

In order to verify the ZYZ rotations, we performed a proof-of-principle experiment with cold atomic qubits and as-programmed femtosecond laser pulses (see Fig.~\ref{Fig3}).  
The detail of our laser experimental setup is described in our previous work~\cite{LeePRA2016, SongPRA2016}. Briefly, we used amplified optical pulses from a Ti:sapphire mode-locked laser. Initial pulses were produced at a repetition rate of 1~kHz from the laser, wavelength-centered at the resonance wavelength 795~nm of the rubidium transition from 5S$_{1/2}$ to 5P$_{1/2}$. The spectral bandwidth was 2.5~THz in Gaussian width, equivalent to a pulse duration of 212~fs (FWHM) for a transform-limited (TL) Gaussian pulse. The pulses were then shaped with an acousto-optic pulse programming device (AOPDF, Dazzler from Fastlite)~\cite{AOPDF}.  
The two-level system was formed with the ground and excited states, $|g\rangle=5$S$_{1/2}$ and $|e\rangle=5$P$_{1/2}$, of atomic rubidium ($^{87}$Rb) and the atoms were held in a magneto-optical trap~\cite{LimSR2014}. The inhomogeneity of the laser-atom interaction~\cite{LeeOL2015}, due to the spatial intensity profile of the laser, was minimized by reducing the size of the atom cloud 2.3 times smaller than the the laser beam. The size of the atom cloud was 250~$\mu$m (FWHM).

\begin{figure*}
\centerline{\includegraphics[width=0.8\textwidth]{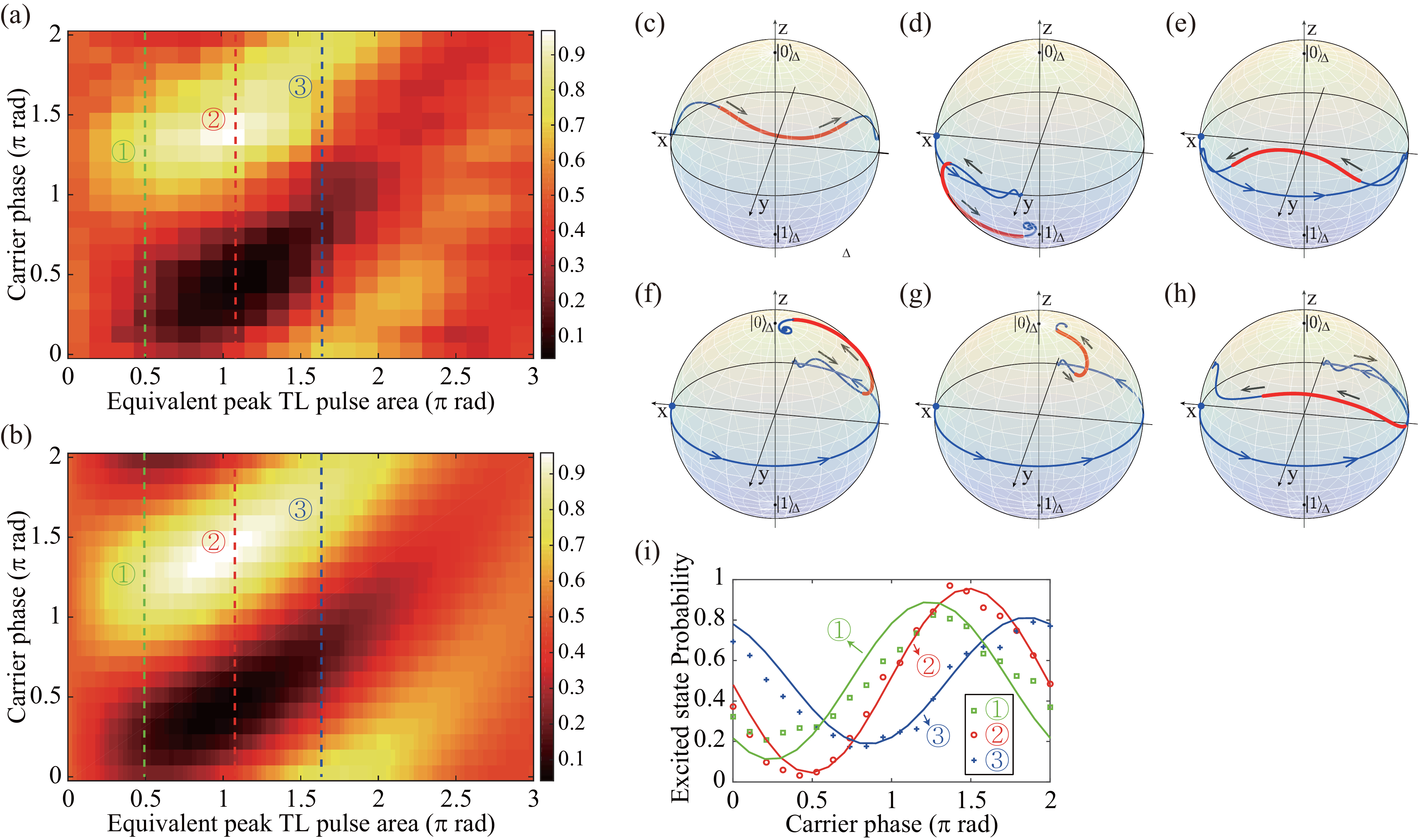}}
    \caption{(Color online) (a) Measured excitation probability $P_e(\Theta, \varphi, \Phi_1)$, of atoms initially in $|\psi_{\rm init}\rangle = (|g\rangle_{\omega_0} + |e\rangle_{\omega_0})/\sqrt{2}$, probed as a function of the equivalent TL pulse-area $\mathcal{A}$ and the carrier phase $\varphi$, where the chirped pulse with a temporal hole is defined in Eq.~\eqref{efieldhole} with $k=0.7$ and $\tau_h=0.4\tau$. (b) The corresponding TDSE calculation. (c-h) Bloch vector dynamics (in the ``detuning'' interaction picture) at selected points: (c) ($\varphi$, $\mathcal{A}$)=(0, $\pi$), (d) ($\pi/2$, $\pi$), (e) ($\pi$, $\pi$), (f) ($3\pi/2$, $\pi$), (g) ($3\pi/2$, $\pi/2$), (h) ($3\pi/2$, $3\pi/2$). (i) Comparison between the experimental results and calculation along the three dashed lines in (a) and (b).}
    \label{Fig4}
\end{figure*}

The control experiment was conducted in three steps: initialization, qubit rotation, and detection. The atoms were first excited by a $\pi/2$-area pulse to initialize the atoms in the superposition state $|\psi_{\rm init}\rangle$ defined in Eq.~\eqref{eq.init}. Then, the chirped pulse with a temporal hole rotated the state.
Lastly, atoms in the excited state were detected through ionization, using a frequency-doubled split-off of an un-shaped laser pulse and a micro-channel plate (MCP) detector. 

The laser pulses for the initialization and qubit rotation were programmed by the AOPDF. In the frequency domain, the combined field is given by
\begin{eqnarray}
\Tilde{E}(\omega) = \Tilde{E}_{\rm init}(\omega)+\Tilde{E}_{\rm rot}(\omega)e^{i\varphi},
\label{shapingeq}
\end{eqnarray}
where $\Tilde{E}_{\rm init}(\omega)$ is the $\pi/2$-area pulse, $\Tilde{E}_{\rm rot}(\omega)$ is the chirped pulse with a temporal hole, and $\varphi$ is the relative phase between them. The total energy of these two pulses was up to 20 $\mu$J and the energy of each pulse was pre-calibrated through cross-correlation measurements. The chirp parameter for the control pulse was fixed at $\alpha=8.15$~rad/ps$^2$, which corresponds to the frequency chirp of 60,000~fs$^2$ in the spectral domain. 


Figure~\ref{Fig4} shows a comparison between experimental and theoretical results. When atoms, in the initial superposition state $|\psi_{\rm init}\rangle$ in Eq.~\eqref{eq.init}, undergo the rotation in Eq.~\eqref{eq.rrr}, the excited-state probability is given by
\begin{eqnarray}
P_e (\Theta, \varphi, \Phi_1) 
&=& |\langle e|\mathcal{R}_{\hat{z}}(\Phi_2)\mathcal{R}_{\hat{y}}(\Theta)\mathcal{R}_{\hat{z}} (\Phi_1+\varphi)|\psi_{\rm init}\rangle|^2 \nonumber \\
&=& \frac{1}{2}\left[1 - \sin \Theta \cos(\Phi_1+\varphi)\right]
\label{eqevol2}.
\end{eqnarray}
The resulting behavior of $P_e$ is an oscillatory function, of which the amplitude and phase are determined by $\Theta$ and $\Phi_1+\varphi$. In Fig.~\ref{Fig4}(a), the measured probability is plotted as a function of the equivalent (peak) TL pulse-area $\mathcal{A}$ and the carrier phase $\varphi$. The result strongly agrees with the calculation in Fig.~\ref{Fig4}(b), performed with the corresponding time-domain Schr\"{o}dinger equation (TDSE). 
Each point in Figs.~\ref{Fig4}(a) and \ref{Fig4}(b) corresponds to a distinct Bloch vector evolution. A few characteristic trajectories (in the ``detuning'' interaction picture) are shown in Figs.~\ref{Fig4}(c,d,$\cdots$,h) (see the figure caption for more detail). 

Along the dashed lines in Figs.~\ref{Fig4}(a) and \ref{Fig4}(b), data points are extracted and compared in Fig.~\ref{Fig4}(i), where the excited-state probabilities, $P_e(\varphi|\Theta,\Phi_1)$, are plotted as a function of $\varphi$ at fixed $\Theta$ and $\Phi_1$. The change of the peak oscillation point in Fig.~\ref{Fig4}(i) is related to the  $E_0$-dependence of $\Phi_1$ as in Eq.~\eqref{eqevol2}; $\Phi_1$ is a monotonically decreasing function of $E_0$, so the peaks in Fig.~\ref{Fig4}(i) shift to the upper right corner as $E_0$ increases.  Also, the change in the oscillation amplitude is related to the $E_0$-dependence of $\Theta$. As the electric-field amplitude $E_0$ increases, so does the rotation angle $\Theta$ of the Y-rotation; however, it is up to a certain maximum $E_0$, at above of which the dynamics involved with the hole gradually becomes adiabatic. Such behavior of $\Theta$ is clearly demonstrated in Fig.~\ref{Fig4}(i), where the oscillation amplitude given by $\sin\Theta$ in Eq.~\eqref{eqevol2} reaches maximal, along the line marked by \raisebox{.5pt}{\textcircled{\raisebox{-.7pt} {2}}}, and decreases as $E_0$ increases. Therefore, the expected behaviors of $\Phi_1$ and $\Theta$ in Eq.~\eqref{eqevol2} are clearly observed in the experimental results.

\section{conclusion}

In summary, we proposed and demonstrated the use of hybrid adiabatic and non-adiabatic interaction for  single-laser-pulse implementation of arbitrary qubit rotations. The chirped optical pulse with a temporal hole induced  ZYZ-decomposed rotations of atomic qubits all at once, in which the temporal hole caused a non-adiabatic evolution amid an otherwise monotonic adiabatic evolution due to the chirped pulse. The proof-of-principle experimental verification of the given laser-atom interaction was performed with programmed femtosecond laser pulses and cold atoms. The result suggests that laser pulse-shape programming may be useful in quantum computation through concatenating gate operations in a quantum circuit.

\begin{acknowledgements}
This research was supported by Samsung Science and Technology Foundation [SSTF-BA1301-12]. Authors thank Adam Massey and Chansuk Park for fruitful discussions. 
\end{acknowledgements}

\appendix
\section{Chirped pulses in frequency and time domains}
A linearly chirped pulse is defined with a second-order phase in the spectral domain, which can be written as
\begin{equation}
\widetilde{E}_{\rm chirp}(\omega) =  \frac{E_0}{\sqrt{2}\Delta\omega}\exp\Big[{-\frac{(\omega-\omega_0)^2}{\Delta \omega^2}} -\frac{i c_2}{2}(\omega-\omega_0)^2\Big],
\label{ap_E_omega}
\end{equation}
where a Gaussian pulse with amplitude $E_0$ and frequency chirp $c_2$ is assumed and the frequency is centered at the resonance $\omega_0$ of the two-level system. Then, the time-domain electric field is given by
\begin{eqnarray}
E_{\rm chirp}(t) = E_0\sqrt{\frac{\tau_0}{\tau}}e^{-{t^2}/{\tau^2}} \cos{[(\omega_0 +\alpha t )t + \varphi]} 
\label{ap_e-field},
\end{eqnarray}
where $\varphi =-\tan^{-1}(2c_2/\tau_0^2)/2$ is the phase, $\tau_0 = {2}/{\Delta\omega}$ the transform-limited (TL) pulse width,  $\tau=\sqrt{\tau_0^2+4c_2^2/\tau_0^2}$ the chirped pulse width, and $\alpha = {2c_2}/(\tau_{0}^4+4c_2^2)$ the chirp parameter. 

\section{Hamiltonian transformation}
The dynamics of a two-level system interacting with a shaped chirped pulse is governed by the Hamiltonian 
\begin{eqnarray}
H = \left[ \begin{array}{cc}
-\hbar\omega_0/2 & \mu E(t) \\
\mu E(t) & \hbar\omega_0/2
\end{array} \right],
\label{ap_Hamiltonian1}
\end{eqnarray}
where the two base vectors are defined as $|g\rangle$ and $|e\rangle$. Being transformed to the ``field'' interaction picture (with respect to the instantaneous laser frequency $\omega_L(t)=\omega_0+2\alpha t$), the Hamiltonian $H$ becomes
\begin{eqnarray}
H_{\omega_L} = \frac{\hbar}{2} \left[ \begin{array}{cc}
-\Delta(t) & \Omega(t) \\
\Omega(t) & \Delta(t)
\end{array} \right], 
\label{ap_Hamiltonian2}
\end{eqnarray}
after the rotating wave approximation, where $\Delta(t)=\omega_0-\omega_L(t)=-2\alpha t$ is the instantaneous detuning and $\Omega(t)$ is the Rabi frequency. The transformation matrix from $H$ to $H_{\omega_L}$ is given by $\mathcal{T}_{\omega_L}=\exp(i\int_0^t T_{\omega_L} (t') dt'/\hbar)$ with
\begin{eqnarray}
T_{\omega_L} = \frac{\hbar}{2} \left[ \begin{array}{cc}
-(\omega_0 t+ \alpha t^2) & 0 \\
0 & (\omega_0 t+ \alpha t^2)
\end{array} \right]
\end{eqnarray}
where the base vectors in the ``field'' interaction picture are $|g\rangle_{\omega_L} = \mathcal{T}_{\omega_L}|g\rangle$ and $|e\rangle_{\omega_L} = \mathcal{T}_{\omega_L}|e\rangle$.

Chirp pulses induce adiabatic evolution, which is a Z-rotation in the adiabatic basis. The adiabatic base vectors are given by
\begin{eqnarray}
|0(t)\rangle_A &=& \cos\vartheta(t)|g\rangle_{\omega_L} -\sin\vartheta(t)|e\rangle_{\omega_L}, \nonumber \\
|1(t)\rangle_A &=& \sin\vartheta(t)|g\rangle_{\omega_L} +\cos\vartheta(t)|e\rangle_{\omega_L}
\label{ap_adiaeq2},
\end{eqnarray}
where the eigenvalues are
\begin{equation}
\frac{\hbar}{2}\lambda_\pm(t) = \pm \frac{\hbar}{2}\sqrt{\Omega^2(t)+\Delta^2(t)}
\label{ap_adiaeq1}
\end{equation}
and the mixing angle $\vartheta(t)$ is 
\begin{equation}
\vartheta(t) = \frac{1}{2} \tan^{-1} \frac{\Omega(t)}{\Delta(t)} \quad \rm{for} \quad 0 \le\vartheta(t)\le\frac{\pi}{2}.
\label{ap_mixingangle}
\end{equation}
The state in the adiabatic basis is given by $|\psi(t)\rangle_A=R(\vartheta(t))|\psi(t)\rangle_{\omega_L}$, where $|\psi(t)\rangle_{\omega_L} = \mathcal{T}_{\omega_L}|\psi(t)\rangle$ and $R(\vartheta(t))$ is the adiabatic transform matrix defined as
\begin{equation}
R(\vartheta(t))=\left[ 
\begin{array}{cc}
\cos\vartheta(t) & -\sin\vartheta(t) \\
\sin\vartheta(t) & \cos\vartheta(t)
\end{array} 
\right].
\label{ap_adiarot}
\end{equation}
The Schr\"odinger equation is then given in the adiabatic basis $\{|0(t)\rangle_{A}, |1(t)\rangle_{A}\}$ by
\begin{eqnarray}
i\hbar \frac{d }{dt} |\psi(t)\rangle_A = \left( R H_{\omega_L} R^{-1}  +  i\hbar \dot{R} R^{-1}  \right) |\psi(t)\rangle_A,
\label{ap_Seqadia}
\end{eqnarray}
and the adiabatic Hamiltonian is 
\begin{equation}
H_A
= \frac{\hbar}{2} \left[
 \begin{array}{cc}
\lambda_- & -2i\dot{\vartheta}  \\
2i\dot{\vartheta}  & \lambda_+
\label{ap_adiaH1}
\end{array}
\right],
\end{equation}
where $2\dot{\vartheta}$ in the off-diagonal term is the ``non-adiabatic coupling" given by
\begin{equation}
2\dot{\vartheta} = \frac{|\dot{\Omega}(t)\Delta(t)-\Omega(t)\dot{\Delta}(t)|}{\Delta^2(t)+\Omega^2(t)}
\label{ap_diffmix}.
\end{equation} 

With the adiabatic Hamiltonian $H_A$, the phase of the state diverges at $t\to\pm\infty$, because of the detuning. To remove this phase before and after the pulse duration, we perform an additional transform $\mathcal{T}_{\Delta}=\exp\Big({i} \int_{0}^{t} T_\Delta(t') dt'/{\hbar} \Big)$ with 
\begin{equation}
T_\Delta = \frac{\hbar}{2}\left[ \begin{array}{cc}
-|\Delta(t)| & 0  \\
0 & |\Delta(t)|
\end{array} \right].
\end{equation}
The resulting Hamiltonian in this ``detuning'' interaction picture, also in Eq.~(3), is given by
\begin{equation}
H_{\Delta}
= \frac{\hbar}{2} \left[
 \begin{array}{cc}
-\Delta_F(t) & \Omega_F(t)  \\
\Omega_F^{*}(t)  & \Delta_F(t)
\end{array} 
\right],
\label{ap_adiaH2}
\end{equation}
where the modified detuning and Rabi frequency are
\begin{eqnarray}
\Delta_F(t) &=& \sqrt{\Delta^2(t)+\Omega^2(t)}-|\Delta(t)|, \nonumber \\
\Omega_F(t) &=& -2i\dot{\vartheta}e^{-i|\Delta(t)|/2}
\label{ap_adiaH3},
\end{eqnarray}
and the base vectors are defined by $|0(t)\rangle_{\Delta} = \mathcal{T}_\Delta|0(t)\rangle_A$ and $|1(t)\rangle_{\Delta} = \mathcal{T}_\Delta|1(t)\rangle_A$.

On the other hand, the conventional ``atomic'' interaction picture uses the transformation, given by $\mathcal{T}_{\omega_0}=\exp(i\int_0^t T_{\omega_0} (t') dt'/\hbar)$ with
\begin{eqnarray}
T_{\omega_0}
= \frac{\hbar}{2} \left[
 \begin{array}{cc}
-\omega_0 & 0  \\
0 & \omega_0 
\end{array} 
\right]
\label{ap_atomicint1}
\end{eqnarray}
to remove the phase factor associated with the atomic energy splitting $\omega_0$. In this representation (the ``atomic'' interaction picture), the base vectors are given by $|g\rangle_{\omega_0} = \mathcal{T}_{\omega_0}|g\rangle$ and $|e\rangle_{\omega_0} = \mathcal{T}_{\omega_0}|e\rangle$.


\begin{thebibliography}{1}
\bibitem{arbitrary} The rotational axis and angle of a ZYZ-decomposed arbitrary rotation are respectively given by $\hat{n} = ( \sin\frac{\Theta}{2}\sin\Phi_+, \sin\frac{\Theta}{2}\cos\Phi_-, \cos\frac{\Theta}{2}\sin\Phi_+) / \sin \frac{\phi}{2} $ and $\phi=2\cos^{-1} (\cos\frac{\Theta}{2}\cos\Phi_+)$, where $\Phi_\pm=(\Phi_1\pm\Phi_2)/2$.
\end{thebibliography}
\end{document}